\documentclass[prb,
twocolumn,
superscriptaddress,showpacs,amsmath,amssymb]{revtex4}

\begin{document}

\title{From black hole to white hole  via the intermediate static state}

\author{G.E.~Volovik}
\affiliation{Low Temperature Laboratory, Aalto University,  P.O. Box 15100, FI-00076 Aalto, Finland}
\affiliation{Landau Institute for Theoretical Physics, acad. Semyonov av., 1a, 142432,
Chernogolovka, Russia}

\date{\today}

\begin{abstract}
We discuss the macroscopic quantum tunneling from the black hole to the white hole of the same mass. Previous calculations in Ref.[1] demonstrated that the probability of the tunneling is $p \propto \exp(-2S_\text{BH})$, where $S_\text{BH}$ is the entropy  of the Schwarzschild black hole. This in particular suggests that  the entropy of the white hole is with minus sign the entropy of the black hole, $S_\text{WH}(M)=- S_\text{BH}(M)= - A/(4G)$. Here we use a different way of calculations. We consider three different types of the hole objects: black hole, white hole and the fully static intermediate state. The probability of tunneling transitions between these three states is found using singularities in the coordinate transformations between these objects. The black and white holes are described by the Painleve-Gullstrand coordinates with opposite shift vectors, while the intermediate state is described by the static Schwarzschild coordinates. 
The singularities in the coordinate transformations lead to the imaginary part in the action, which determines the tunneling exponent. For the white hole the same negative entropy is obtained, while the intermediate state -- the fully static hole --  has zero entropy. This procedure is extended to the Reissner-Nordstr\"om black hole and to its white and static partners, and also to the entropy and temperature of the de Sitter Universe.
\end{abstract}
\pacs{
}

\maketitle

\section{Introduction}

In Ref. \cite{Volovik2020a} the quantum tunneling from black hole and to white hole with the same mass $M$ was considered.  The probability of such tunneling was $\exp(-2S_\text{BH})$, where $S_\text{BH}$ is the black hole entropy. The quantum transition can be considered as thermodynamic fluctuation determined by the difference in the entropy of the final and initial states, 
$\exp(S_\text{WH} -S_\text{BH})$, from which it follows that the entropy of the white hole (with the same mass $M$  as the black hole) is negative, $S_\text{WH}(M)=- S_\text{BH}(M)$. Also from the thermodynamic potential it follows that the white hole formed from the black hole by quantum tunneling has negative temperature, $T_\text{WH}(M)=- T_\text{BH}(M)$.

The negative entropy is rather unusual phenomenon. That is why we consider the calculation of the quantum tunneling from black hole  to white hole using different approach. In Ref. \cite{Volovik2020a} the tunneling exponent was calculated using the path in the complex plane of the varying gravitational coupling (the inverse Newton "cosntant"). 
Here we consider the intermediate state between the black hole and white hole, which is represented by the fully static metric, and calculate the transition probability from black hole to static hole and then from static hole to the white hole. We shall use the path on the complex coordinate plane, which connects the static coordinates with the stationary but not static coordinates of the black or white holes. As a result the equation $S_\text{WH}(M)=- S_\text{BH}(M)$ is confirmed, while for the intermediate symmetric static hole the entropy is zero.

\section{Schwarzschild holes: black, white and fully static}
\label{SchwarzschildHoles}

\subsection{Fully static hole}

Let us start with Schwarzschild black hole.
The black and white holes have stationary but not static metric ($c=1$), which in Painleve-Gullstrand\cite{Painleve,Gullstrand}  (PG) coordinate system have the form:
 \begin{equation}
ds^2= - dt^2(1-{\bf v}^2) - 2dt\, d{\bf r}\cdot {\bf v} + d{\bf r}^2 \,.
\label{PGmetric}
\end{equation}
Here the shift vector $v_i({\bf r})=g_{0i}({\bf r})$ is the velocity of the free-falling observer, which crosses the horizon. 
For the Schwartzschild black hole one has
\begin{equation}
{\bf v} ({\bf r})=\mp \hat{\bf r} \sqrt{\frac{r_\text{H}}{r}}\,,
\label{velocity}
\end{equation}
where $r_\text{H}$ is the radius of the horizon; the minus sign corresponds to the black hole and the plus sign describes the white hole. In both states the time reversal symmetry is violated, since $T{\bf v}= - {\bf v}$, and thus under time reversal operation the black hole transforms to the white hole.

So, the black and white holes represent two degenerate states, which correspond to the spontaneously broken time reversal symmetry.
This suggests that there can be also the intermediate state, in which the time reversal symmetry is not violated, i.e. the fully static configuration.
The role of such static state can be played by the configuration described by the fully static metric:
\begin{equation}
ds^2=- \left(  1- \frac{r_\text{H}}{r} \right)dt^2 + \frac{dr^2}{\left(  1- \frac{r_\text{H}}{r} \right)} + r^2 d\Omega^2
\,.
\label{StaticMetric}
\end{equation}
This metric has coordinate singularity at the horizon, but it is invariant under time reversal.
Such static hole has the same mass as the black hole state. We consider this state as the intermediate state in the quantum tunneling from the black hole (BH) to white hole (WH). The probability of such transition  via the static state has probability:
\begin{equation}
P_{\text{BH}\rightarrow \text{WH}} = P_{\text{BH}\rightarrow \text{static}}P_{\text{static}\rightarrow \text{WH}}\,,
\label{transition1}
\end{equation}
with $P_{\text{BH}\rightarrow \text{static}}=P_{\text{static}\rightarrow \text{WH}}$ as follows from symmetry between black and white hole. 

\subsection{Tunneling from black hole to static hole}

Let us calculate $P_{\text{BH}\rightarrow \text{static}}$ using the approach of the semiclassical quantum tunneling, which originally has been used for the calculation of the Hawking radiation.\cite{Wilczek2000,Srinivasan1999,Volovik1999,Akhmedov2006,Vanzo2011} The transformation from static hole with metric (\ref{StaticMetric}) to the black hole with PG metric (\ref{PGmetric}) is obtained by the singular coordinate transformation:
\begin{equation}
dt \rightarrow dt +dr \frac{v}{1-v^2} \,.
\label{CoordinateTransformation}
\end{equation}
Such singularity in the coordinate transformation drastically changes the action $\int M dt$, which demonstrates that  the static hole and the black hole are absolutely different objects. However, in quantum mechanics these two states can be connected by path on the complex coordinate plane, which can be considered as the path along the trajectory describing the macroscopic quantum tunneling between the two hole objects.
The macroscopic quantum tunneling is the tunneling of the macroscopic objects, such as vortex in superfluid \cite{Volovik1972}
and even the whole Universe.

The probability of the quantum tunneling from the black hole to its static partner with the same mass $M$ is given by the imaginary part in the action on this path:
\begin{eqnarray}
 P_{\text{BH}\rightarrow \text{static}} = \exp{ \left(-2{\rm Im} \int M dt \right)} = 
 \label{ImaginaryAction1}
 \\
 = \exp{ \left(-2{\rm Im} \int dr \frac{Mv}{1-v^2}  \right)} =
 \label{ImaginaryAction2}
 \\
 =  \exp{ \left(-2\pi M r_\text{H}\right)} =  \exp{ \left(-4\pi M^2G\right)} 
 =  \exp{ \left(-S_\text{BH} \right)} \,.
\label{ImaginaryAction}
\end{eqnarray}
In the transition from Eq.(\ref{ImaginaryAction1}) to  Eq.(\ref{ImaginaryAction2})  the transformation in Eq.(\ref{CoordinateTransformation}) was used.

Note that in the above derivation the conventional Hawking radiation of particles from the black hole is not taken into account. Here only one channel of the decay of the black hole is considered: the macroscopic tunneling to static hole and then to white hole. The combined many-channel decay is for further consideration in future. In the absence of Hawking radiation, the mass $M$ of the hole object is conserved, and thus the action is simply $\int M dt$, which acquires the imaginary part from the singularity in the coordinate transformation. 

From Eq.(\ref{ImaginaryAction}) it follows that the probability of the macroscopic quantum tunneling from black hole to static state is fully determined by the black hole entropy $S_\text{BH}$. On the other hand the quantum tunneling transition can be considered  as quantum fluctuation and thus it is determined by the difference in the entropy of the initial and final states:\cite{Landau_Lifshitz}
\begin{equation}
P_{\text{BH}\rightarrow \text{static}} =\exp{ \left(S_\text{static} -S_\text{BH} \right)} \,.
\label{transition5}
\end{equation}
Then from Eqs.(\ref{ImaginaryAction}) and (\ref{transition5})  it follows that the entropy of the static hole is zero, $S_\text{static}=0$. This is rather natural, because here the Schwarzschild  static solution of Einstein equations described by metric (\ref{StaticMetric}) is considered not as the black hole, but as the intermediate state between the black and white holes, which has two independent static patches. The intermediate state has no Hawking radiation, its horizon represents  the surface of infinite red shift. This static hole has no absorption or emission of particles, instead there is a reflection of particles from such horizon on both sides of the coordinate singularity. This state should have  zero temperature.

Here we note that in general relativity not all coordinate transformations are allowed. There are some singular transformations, which change the state of the system, i.e. they transfer the initial state  to the final state, which is physically different from the initial state. The transformation from the PG coordinates for black hole to the PG coordinates for white hole belongs to this class. BH and WH are physically different systems, with different entropies and temperatures. The other situations, when two frames are not thermodynamically equivalent, have been discussed in Refs.\cite{Odintsov2020,Odintsov2017}.

The same concerns the singular transformation from the PG coordinates for black hole to the fully static Schwarzschild coordinates. This transformation produces the intermediate static state, which is physically not the BH and not the WH. This does not contradict to the consideration of the black hole in Schwarzschild coordinates. When the real black hole is considered using Schwarzschild coordinates, the singularity is usually avoided by some tricks, which correspond either to the black hole  behaviour at the horizon (or  to the white hole behaviour at the horizon if the white hole is considered). But here we consider the truly static state as the real solution of Einstein equations, which serves as the intermediate state between the black hole and the white hole. Also the particles propagating at this static spacetime are also considered without ascribing the black or white hole features to the horizon. 
This means that all the three states (black, white and intermediate static), which cannot be obtained from each other by the regular coordinate transformations, are physically different.

\subsection{Tunneling from black hole to white hole}

Let us now consider  the quantum tunneling from black hole to white hole. From Eqs. (\ref{transition1})  and (\ref{ImaginaryAction}) it follows that the probability of the tunneling is:
\begin{equation}
P_{\text{BH}\rightarrow \text{WH}} = P^2_{\text{BH}\rightarrow \text{static}} =  \exp{ \left(-8\pi M^2G\right)} =\exp{ \left(-2S_\text{BH} \right)} .
\label{transition3}
\end{equation}
The same result (\ref{transition3}) can be obtained using the direct transformation from the black hole to the white hole:
\begin{equation}
dt \rightarrow dt +2dr \frac{v}{1-v^2} \,.
\label{CoordinateTransformationBHWH}
\end{equation}

Equation of the type $\exp{ \left(-\gamma M^2G\right)}$ for the probability of the black hole to white hole tunneling is expected on dimensional grounds, see e.g. Ref.\cite{Rovelli2018} (see also Refs.\cite{Barcelo2014,Barcelo2017,Rovelli2019,Rovelli2018b,Uzan2020,Uzan2020b,Uzan2020c,Uzan2020d,Bodendorfer2019} for the black hole to white hole transitions).
However, the parameter $\gamma=8\pi$ in Eq.(\ref{transition3}) leads to rather unexpected result, see Sec.\ref{NegEntropy}.

\subsection{Negative entropy of white hole}
\label{NegEntropy}

Due to connection between quantum fluctuations and the difference in entropy between the initial and final states\cite{Landau_Lifshitz} this probability has the form:
\begin{equation}
P_{\text{BH}\rightarrow \text{WH}} =\exp{ \left(S_\text{WH} -S_\text{BH} \right)} \,.
\label{transition4}
\end{equation}
Then from Eqs.(\ref{transition3}) and (\ref{transition4}) one obtains the result of Ref. \cite{Volovik2020a}, that the white hole entropy is with minus sign  the entropy of the black hole  with the same mass,  $S_\text{WH}(M)=- S_\text{BH}(M)$:
\begin{equation}
S_\text{WH}= S_\text{BH}-2S_\text{BH}= -S_\text{BH}
 \,.
\label{WhitHoleEntropy}
\end{equation}
The black hole states with negative entropy have been considered in Ref.\cite{Odintsov2002}, where it has been suggested that appearance of negative entropy may indicate a new type instability, see also Ref.\cite{Odintsov2020}.
Such super-low entropy of white hole can be also seen as an example of a memory effect discussed in Ref.\cite{Rovelli2020}, i.e.
the entropy of the white hole is negative, since this state remembers that it is formed from the black hole by quantum tunneling.

The negative entropy also results in the negative temperature of the white hole\cite{Volovik2020a}
(see also Refs.\cite{Lounasmaa1997,Baldovin2021,VolovikNegT} on negative temperature in condensed matter and relativistic physics). 

\section{Reissner-Nordstr\"om black, white and static holes}

\subsection{Modified Painleve-Gullstrand coordinates}
\label{ModifiedPG}

Let us apply the same procedure for the Reissner-Nordstr\"om (RN) black hole, which has two horizons.
The static state of the RN hole has the metric with two singularities, on the outer horizon at $r_+$  and on the inner horizon at $r_-$:
 \begin{equation}
ds^2= - dt^2\frac{(r-r_-)(r-r_+)}{r^2} + dr^2\frac{r^2}{(r-r_-)(r-r_+)} +r^2 d\Omega^2.
\label{staticRN}
\end{equation}
The positions of horizons  are expressed in terms of the mass $M$ and charge $Q$ of the hole  ($G=c=1$):
 \begin{equation}
r_+r_-= Q^2\,\,, \,\, r_+ + r_-= 2M\,.
\label{r+r-}
\end{equation}

The coordinate transformation:\cite{Volovik2003}
 \begin{equation}
dt \rightarrow dt \pm fdr  \,\,,\, f=\frac{\sqrt{2Mr}}{\sqrt{r^2 + Q^2}}\frac{r^2}{(r-r_-)(r-r_+)}\,,
\label{CoordinateTrnaformationRN}
\end{equation}
leads to the black hole and white hole states without singularities at horizons:
 \begin{equation}
ds^2= - dt^2\left( 1 +\frac{Q^2}{r^2}\right)+ \frac{1}{1 +\frac{Q^2}{r^2}} (dr\pm v dt)^2+r^2 d\Omega^2\,,
\label{PG_RN}
\end{equation}
where the shift velocity is:
 \begin{equation}
v^2 = \frac{2M}{r} \left( 1 +\frac{Q^2}{r^2}\right)\,.
\label{velocity_RN}
\end{equation}
This form represents the extension of the PG coordinates, in which the shift velocity is real for all $r$. This is distinct from the non-modified PG coordinates,\cite{Hamilton2008,Zubkov2019} where the shift velocity becomes imaginary for $r<r_0= r_+r_-/(r_+ +r_-)$.

\subsection{Quantum tunneling and entropy}
\label{TunnelingAndEntropy}

The regularity of the shift velocity and of the whole metric demonstrates that Eq. (\ref{PG_RN}) fully describes the  Reissner-Nordstr\"om black and white holes, and thus  one can use the coordinate transformation (\ref{CoordinateTrnaformationRN}) for the calculation of the quantum transitions between the hole objects.

The probability of the quantum tunneling from the RN black hole to its static RN partner in Eq.(\ref{staticRN}) with the same mass $M$ and charge $Q$ is determined now by two singularities (at two horizons). Using Eq.(\ref{CoordinateTrnaformationRN}) for coordinate transformation one obtains
\begin{eqnarray}
 P_{\text{BH-RN}\rightarrow \text{static-RN}}  =  \exp{ \left(-2\,{\rm Im} \int M dt \right)} =
 \label{P_RN1}
 \\
 = \exp{ \left(-2M\,{\rm Im} \int f(r)dr  \right)} = 
 \label{P_RN2}
 \\
 = \exp{ \left(-\pi (r_+ + r_-)^2/G\right)} =   \exp{ \left(-4\pi M^2G\right)} \,.
\label{P_RN3}
 \end{eqnarray}

The probability of the tunneling appears to be the same as for the Schwarzschild black hole with the same mass $M$. Under natural assumption that the entropy of the static hole is zero, the Eq.(\ref{P_RN3}) suggests that the entropy of the RN black hole with both horizons taken into account is the same as the entropy of the Schwarzschild black hole with the same mass, i.e. it does not depend on charge, 
\begin{equation}
S_\text{BH-RN}(M,Q)=S_\text{BH}(M,0)  \,.
\label{entropy_RN}
\end{equation}
This demonstrates that the factor $(1+Q^2/r^2)$ in Eqs.(\ref{PG_RN}) and (\ref{velocity_RN})  is irrelevant for the entropy and can be removed by adiabatic process. This probably has something to do with the discussed controversies concerning the near extremal black holes.\cite{Page2019} The Hawking radiation in the system with two horizons is also modified. Instead of the single tunneling event from the outer horizon one should consider the process of co-tunneling -- the simultaneous tunneling from two horizons. The similar co-tunneling was considered for Hawking radiation of particles with mass $m<E$.\cite{Jannes2011}

As in the case of the Schwarzschild holes in Sec.\ref{SchwarzschildHoles}, the  consideration of the further tunneling -- from the static RN state to the RN white hole state -- provides the negative entropy for the Reissner-Nordstr\"om white hole,
 \begin{equation}
S_\text{WH-RN}(M,Q)=- S_\text{BH-RN}(M,Q)
 \,.
\label{entropy_RN2}
\end{equation}
 
 \section{Entropy and temperature of de Sitter Universe}

Calculation of energy, entropy and laws of thermodynamics in the case of de Sitter spacetime is still an open problem, see e.g. Ref. \cite{Padmanabhan2002}. Let us apply the same procedure of singular coordinate transformation for the calculations of the de Sitter entropy. The corresponding PG metric is given by Eq.(\ref{PGmetric}) with ${\bf v}( {\bf r})=\pm H {\bf r}$, where $H$ is the
Hubble parameter in the de Sitter universe, while the static metric is:
\begin{equation}
ds^2=- \left( 1- H^2r^2\right)dt^2 + \frac{dr^2}{1- H^2r^2} + r^2 d\Omega^2
\,.
\label{dSStaticMetric}
\end{equation}
The corresponding coordinate transformation between the expanding and static metrics is given by Eq.(\ref{CoordinateTransformation}).
The probability of the quantum tunneling from the de Sitter to its static partner with the same energy $E$ is given by the imaginary part in the action:
\begin{eqnarray}
 P_{\text{dS}\rightarrow \text{static}} = \exp{ \left(-2{\rm Im} \int E dt \right)} = 
 \label{ImaginaryAction1dS}
 \\
 = \exp{ \left(-2{\rm Im} \int dr \frac{Ev}{1-v^2}  \right)} =
 \label{ImaginaryAction2dS}
  \\
 = \exp{ \left(-2{\rm Im} \int dr \frac{EHr}{1-H^2r^2}  \right)} =
 \label{ImaginaryAction3dS}
 \\
 =  \exp{ \left(-\frac{\pi E}{H}\right)} \,.
\label{ImaginaryActiondS}
\end{eqnarray}

From the connection between the tunneling transition and the difference in the entropy of the initial and final states it follows that  the entropy of the expanding Universe is 
\begin{equation}
S_\text{dS}= \frac{\pi E}{H}
\,.
\label{dSentropy}
\end{equation}
This equation allows us to study thermodynamics of the de Sitter Universe
The main problem is that the proper energy $E$ of the de Sitter Universe is not well determined, and there are different suggestions such as $E=1/(2H)$,  $E=-1/(2H)$ and  $E=-1/H$.\cite{Padmanabhan2002}

Let us assume that the entropy has the conventional form as the quarter of the  area of the cosmological horizon, $S_\text{dS}=A/4=\pi/H^2$. Then from Eq.(\ref{dSentropy})  one obtains:
\begin{equation}
E_\text{dS}=\frac{1}{H} = \frac{8\pi}{3H^3}\Lambda
\,,
\label{dSenergy}
\end{equation}
which corresponds to the vacuum energy in the whole volume of the expanding patch, $V= 8\pi/3H^3$.

This allows us to consider the temperature of the de Sitter Universe. If one determines temperature as simply the variation of energy over entropy, one obtains that the temperature of the de Sitter Universe coincides with the Hawking temperature of the cosmological horizon:
\begin{equation}
T_\text{dS}= \frac{dE}{dS}=\frac{H}{2\pi}=T_\text{H}
\,.
\label{dStemperature}
\end{equation}
However, all this is not so simple since we did not take into account the term $-pdV$, where $p$ is the vacuum pressure, $p=-\Lambda$.

To avoid the pressure term we can vary the entropy in Eq.(\ref{dSentropy}) over $dE$ at fixed $H$, which corresponds to the variation at fixed volume $V$ (this can be done by varying the Newton constant $G$ without  changing $H$).
Then one obtains 
\begin{equation}
T_\text{dS}= \frac{dE}{dS}=\frac{H}{\pi}=2T_\text{H}
\,.
\label{dStemperature2}
\end{equation}
This double Hawking temperature of the de Sitter Universe has been obtained in Ref. \cite{VolovikDouble}. This temperature describes different  thermal processes  in the de Sitter Universe, which are not allowed in the Minkowski vacuum, such as ionization of atoms,  splitting of the composite particle with mass $M$ into two components with $M_1 +M_2>M$, and the decay of massive scalar field.\cite{Bros2008,Bros2010,Jatkar2012}  

\section{Conclusion}

 There are different routes of the transformation of the black hole to the white hole. The typical scenario is the decay of the black hole due to Hawking radiation, which is accompanied by the further tunneling to the white hole with the smaller mass.
\cite{Rovelli2018,Barcelo2014,Barcelo2017,Rovelli2019,Rovelli2018b,Uzan2020,Uzan2020b,Uzan2020c,Uzan2020d,Bodendorfer2019} 
We considered the direct tunneling to the white hole with the conservation of mass. Such process has extremely low probability, when compared to the other processes of the decay of the black hole. It may work only on the latest stage of the decay, i.e. at the end of the Hawking evaporation, when the mass is sufficiently small and the quantum tunneling cannot be neglected.  However, the calculations of the probability of this process allows us to find the entropy of the white hole.

Here we derived this probability using the method of singular coordinate transformations.
The result for the tunneling exponent coincides with that obtained earlier by two different methods used in Ref. \cite{Volovik2020a}. The same approach applied to the de Sitter Universe, reproduces the earlier results \cite{Bros2008,Bros2010,Jatkar2012}  of the double Hawking temperature of the de Sitter Universe.  
 The full agreement with previous calculations confirms the validity of the procedure used in this paper. 
 
  In this procedure it is suggested that the fully static state, which is the intermediate state between the black and white holes,  has zero entropy. Then the singular coordinate transformation from the static state to the non-static but stationary state determines the entropy change and thus the thermodynamics of the stationary state. This demonstrates that some singular coordinate transformations in general relativity do transform the initial state to the physically (thermodynamically) different state. The choice of the coordinate system assumes the choice of the reference frame for thermodynamics. This represents an example of the spontaneously broken symmetry with respect to the general coordinate transformations. While the physical laws are invariant under the coordinate transformations, the degenerate states (different states with the same energy) are transformed to each other.

 Application of this procedure to the black hole and its white and fully static partners shows that these three objects are physically different systems with different thermodynamics. The choice of the zero entropy for the thermodynamics of the intermediate  fully static hole completely determines the entropy and temperature of the black and white holes.

The presented approach also supports the statement\cite{Volovik2020a} that the (anti)symmetry between the black and white holes can be extended to their entropy and temperature.
The Schwarzschild black hole and the Schwarzschild white whole are described by the metrics with opposite shift vectors. The shift vector changes sign under time reversal, which transforms the BH to WH. The absence of the time reversal invariance for each of these holes makes these states non static, but still the metric is stationary (time independent), and thus the entropy and temperature can be well defined.
The Schwarzschild BH and Schwarzschild WH have the opposite entropies  $S_\text{WH}(M)=- S_\text{BH}(M)$ and the opposite Hawking temperatures, $T_\text{WH}(M)=- T_\text{BH}(M)$. For the intermediate static hole the time reversal symmetry is not violated, and this object has zero temperature and zero entropy, $S_\text{static} =T_\text{static} =0$.

This approach can be extended to the other types of the black holes, and correspondingly to their static and white hole partners. Here we considered the extension to the Reissner–Nordstr\"om black hole with two horizons and obtained the similar results: $S_\text{WH}(M)=- S_\text{BH}(M)$ and  $S_\text{static}=0$. However, the mass dependence of entropy of RN black hole $S_\text{BH}(M)$ appeared to be the same as for the Schwarzschild BH with the same mass $M$. This deviates from the area law, which can be ascribed to the contribution of both horizons to entropy.  

It would be interesting to consider the other objects including  the Kerr black and white holes, where time reversal symmetry is violated by rotation. In this case the coordinate transformations produces the singularity in action not only in $\int M\,dt$, but also in $\int J\,d\phi$, where $J$ is angular momentum and $\phi$ is the polar coordinate. The proper coordinate system can be found in Refs.\cite{Doran2000,Hamilton2008}. The recent discussion of the Painleve-Gullstrand forms and their extensions can be found in Refs. \cite{Faraoni2020,Visser2020}. 

The approach has been also applied to the de Sitter Universe. It reproduces the double Hawking temperature of the de Sitter Universe. This can be extended to the black holes in the de Sitter spacetime, such as Reissner-Nordstr\"om-de Sitter black hole.

  {\bf Acknowledgements}. I thank M. Zubkov and A. Zelnikov for discussions, T. Jacobson for criticism and S. Odintsov for attracting my attention to his papers on negative entropy and on the thermodynamically nonequivalent frames. This work has been supported by the European Research Council (ERC) under the European Union's Horizon 2020 research and innovation programme (Grant Agreement No. 694248).


\begin{thebibliography}{999}

 \bibitem{Volovik2020a}
 G.E. Volovik,
Varying Newton constant and black hole to white hole quantum tunneling,
MDPI, Universe {\bf 6}, 133 (2020),
arXiv:2003.10331.

\bibitem{Painleve} 
P. Painlev\'e, 
La m\'ecanique classique et la th\'eorie de la relativit\'e, 
 C. R. Acad. Sci. (Paris) {\bf 173} , 677 (1921).
 
 \bibitem{Gullstrand} 
A. Gullstrand,
 Allgemeine L\"osung des statischen Eink\"orper-problems in der Einsteinschen Gravitations-theorie,
Arkiv. Mat. Astron. Fys. {\bf 16}, 1-15 (1922).

\bibitem{Wilczek2000}
M.K. Parikh and F. Wilczek, 
Hawking radiation as tunneling,
Phys. Rev. Lett. {\bf 85}, 5042 (2000).

\bibitem{Srinivasan1999}
K. Srinivasan and T. Padmanabhan,
Particle production and complex path analysis,
Phys. Rev. D {\bf 60}, 024007 (1999).

 \bibitem{Volovik1999}
G.E. Volovik,  
Simulation of Painleve-Gullstrand black hole in thin $^3$He-A film,  
Pis'ma ZhETF {\bf 69}, 662 -- 668 (1999), JETP Lett.  {\bf 69}, 705 -- 713 (1999); 
gr-qc/9901077.

\bibitem{Akhmedov2006}
E.T. Akhmedov, V. Akhmedova, D. Singleton,
Hawking temperature in the tunneling picture,
Phys. Lett. B {\bf 642}, 124--128 (2006). 

\bibitem{Vanzo2011}
L. Vanzo, G. Acquaviva and R. Di Criscienzo,
Tunnelling methods and Hawking’s radiation: achievements and prospects,
Class. Quantum Grav. {\bf 28}, 183001  (2011).

 \bibitem{Volovik1972}
 G.E. Volovik, 
 Quantum mechanical creation of vortices in superfluids,
 JETP Lett. {\bf 15}, 81 (1972).  

\bibitem{Landau_Lifshitz}
 L.D. Landau  and  E.M. Lifshitz, 
 Course of Theoretical Physics, Volume 5, 
 Statistical Physics.

\bibitem{Odintsov2020}
G.G.L. Nashed, W. El Hanafy, S.D. Odintsov and V.K. Oikonomou,
Thermodynamical correspondence of $f(R)$ gravity in Jordan and Einstein frames,
Int. J. Mod. Phys. D {\bf 29}, 1750154 (2020),
arXiv:1912.03897.

\bibitem{Odintsov2017}
S. Bahamonde, S.D. Odintsov, V.K. Oikonomou, P.V. Tretyakov,
Deceleration versus acceleration universe in different frames of $F(R)$ gravity, 
Phys. Lett. B {\bf 766}, 225--230  (2017).

\bibitem{Rovelli2018}
E. Bianchi, M. Christodoulou, F. D’Ambrosio, H.M. Haggard and C. Rovelli,
White holes as remnants: a surprising scenario for the end of a black hole,
Class. Quantum Grav. {\bf 35},  225003 (2018).

\bibitem{Barcelo2014}
C. Barcelo, R. Carballo-Rubio, L.J. Garay,
Mutiny at the white-hole district,
 Int. J. Mod. Phys. D  {\bf 23},  1442022 (2014).

\bibitem{Barcelo2017}
C. Barcelo, R. Carballo-Rubio, L.J. Garay,
Exponential fading to white of black holes in quantum gravity,
Class. Quantum Grav. {\bf 34}, 105007 (2017).

\bibitem{Rovelli2019}
P.  Martin-Dussaud and C. Rovelli,
Evaporating black-to-white hole,
Class. Quantum Grav. {\bf 36},   245002 (2019).

\bibitem{Rovelli2018b}
C. Rovelli,
Viewpoint: Black hole evolution traced out with loop quantum gravity,
Physics {\bf 11}, 127 (2018).

\bibitem{Uzan2020}
J.B. Achour, J.P. Uzan,
Effective theory of a pulsating Planck star,
arXiv:2001.06153.

\bibitem{Uzan2020b}
J.B. Achour,  S. Brahma and J.P. Uzan,
Bouncing compact objects. Part I. Quantum extension of the Oppenheimer-Snyder collapse,
JCAP 03 (2020) 041.

\bibitem{Uzan2020c}
J.B. Achour and J.P. Uzan,
 Bouncing compact objects. II. Effective theory of a pulsating Planck star,
Phys. Rev. D {\bf 102}, 124041 (2020).

\bibitem{Uzan2020d}
J.B. Achour,  S. Brahma, S. Mukohyama and J.P. Uzan,
Consistent black-to-white hole bounces from matter collapse,
arXiv:2004.1297.

\bibitem{Bodendorfer2019}
N. Bodendorfer, F.M. Mele, J. M\"unch,
Mass and horizon Dirac observables in effective models of quantum black-to-white hole transition,
arXiv:1912.00774.

\bibitem{Odintsov2002}
M. Cvetic, S.Nojiri and S.D. Odintsov,
Black hole thermodynamics and negative entropy in
de Sitter and anti-de Sitter Einstein–Gauss–Bonnet
gravity,
Nuclear Physics B {\bf 628}, 295--330 (2002).

\bibitem{Rovelli2020}
C. Rovelli,
Memory and entropy,
arXiv:2003.06687.

\bibitem{Lounasmaa1997}
A.S. Oja and O.V. Lounasmaa,
Nuclear magnetic ordering in simple metals at positive and negative nanokelvin temperatures,
Rev. Mod. Phys. {\bf 69}, 1-136 (1997).

   \bibitem{VolovikNegT}
G.E. Volovik,
Negative temperature: further extensions,
Pis’ma v ZhETF {\bf 113}, issue 9  (2021),
JETP Lett. {\bf 113}, issue 9  (2021),
arXiv:2104.01013.

\bibitem{Baldovin2021}
M. Baldovin, S. Iubini, R. Livi and A. Vulpiani,
 Statistical mechanics of systems with negative temperature,
 arXiv:2103.12572.
 \bibitem{Volovik2003}
G.E. Volovik, 
What can the quantum liquid say on the brane black hole, the entropy of extremal black hole and the vacuum energy?, 
devoted to the honor of Jacob Bekenstein, Foundations of Physics {\bf 33}, 349--368 (2003);
gr-qc/0301043.

\bibitem{Hamilton2008}
A.J.S. Hamilton and J.P. Lisle, 
The River model of black holes, Am. J. Phys. {\bf 76}, 519 (2008);
[gr-qc/0411060.

\bibitem{Zubkov2019}
M. Lewkowicz and M.A. Zubkov,
Classical limit for Dirac fermions with modified action in the presence of the black hole,
Symmetry {\bf 11}, 1294 (2019),
arXiv:1906.11707.

\bibitem{Page2019}
Shi-Qian Hu, Yen Chin Ong and D.N. Page,
No evidence for violation of the second law in extended black hole thermodynamics,
Phys. Rev D {\bf 100}, 104022 (2019).

\bibitem{Jannes2011}
G. Jannes,
Hawking radiation of $E<m$ massive particles in the tunneling formalism,
JETP Lett. {\bf 94},  18--21  (2011).

\bibitem{Padmanabhan2002}
T. Padmanabhan,
Classical and quantum thermodynamics of horizons in spherically symmetric spacetimes,
Class. Quantum Grav. {\bf 19}, 5387--5408  (2002).

\bibitem{VolovikDouble}
G.E. Volovik,
Double Hawking temperature in de Sitter Universe and cosmological constant problem,
arXiv:2007.05988.

\bibitem{Bros2008} 
J. Bros, H. Epstein, and U. Moschella, 
Lifetime of a massive particle in a de Sitter universe,
JCAP 0802:003 (2008); 
arXiv:hep-th/0612184.

\bibitem{Bros2010} 
J. Bros, H. Epstein, M. Gaudin, U. Moschella and V. Pasquier,
Triangular invariants, three-point functions and particle stability on the de Sitter universe,
Commun. Math. Phys. {\bf 295}, 261--288 (2010).

\bibitem{Jatkar2012} 
D.P. Jatkar, L. Leblond and A. Rajaraman,
Decay of massive fields in de Sitter space,
Phys. Rev. D {\bf 85}, 024047 (2012). 


\bibitem{Doran2000}
C. Doran,
New form of the Kerr solution,
Phys. Rev. D, {\bf 61}, 067503 (2000).

\bibitem{Faraoni2020}
V. Faraoni and G. Vachon,
When Painleve-Gullstrand coordinates fail,
Eur. Phys. J. C {\bf 80}, 771  (2020).

\bibitem{Visser2020}
J. Baines, T. Berry, A. Simpson and M. Visser,
Painleve-Gullstrand form of the Lense-Thirring spacetime,
arXiv:2006.14258


\end{thebibliography}
\end{document}